\documentclass{article}
\usepackage{fleqn,ams}
\usepackage{amssymb}
\usepackage{epsfig}
\usepackage{verbatim}

\author{Yu. G. Ignat'ev\thanks{E-mail: ignatev@kspu.ksu.ras.ru}\\
Department of Geometry, Kazan State Pedagogical University,\\
Mezhlauk 1, Kazan 420021, Russia}
\title{GMSW as a detector of gravitational waves}
\newcommand{\ris}[2]{\refstepcounter{figure}\begin{center}
\begin{tabular}{c}
\parbox{10cm}{\epsfig{file=#1,width=8cm}}\\[12pt]
\parbox{10cm}{\noindent Figure: \thefigure. {\sl #2}}\\
\end{tabular}
\end{center}}

\begin{document}

\maketitle

\begin{abstract}
{\sl The effect of an excitation of a gravitational wave (GW) on
shock waves in a highly magnetized plasma \cite{gmsw}, GMSW, is
studied as an effective means for the detection of GW radiated by
neutron stars. It is shown that there is every reason to identify
the giant impulses of the pulsar NP 0532 with GMSW.}
\end{abstract}

\section{Introduction}
In the History of the •• century Physics, I suppose there is not such a
grave experimental problem (with the exception of the controlled
thermonuclear responses problem) that has been investigated for more than
thirty years by different research groups as the gravitational waves detection
problem (GWDP). Though much means are used to solve it, the GWDP has got
not enough convincing positive result.

The reason for such an unsatisfactory condition in the GWDP, as the author
supposes consists in the mistake of the originally chosen direction of
its solving - in the program of the creation of the gravitational radiation
detectors. The direct detection of GW
can be realized by the tidal GW effect on a nonrelativistic detector
(solid) or due to the relativistic GW effect on a detector which has a
relativistic component (laser ray). In the both cases the GW effect on the
detector (the experimental body displacement or the laser ray deviation)
is propotional to the GW magnitude.
And the expected GW amplitude values of astrophysics sources
are extremely small (see, for example, \cite{torn} and
\cite{rewiev}).
The existing programs of gravitational radiation detection
are calculated generally for astrophysic sources of two types:
1. Supernova; 2. close double systems. In the first case a
GW amplitude which is of the order of $10^{-17} - 10^{-18}$ with the radiation in a
broad frequency spectral range with
the characteristic frequency of the order of $10^3$
sec$^{-1}$ may be expected; in the second case - an amplitude of the
order of $10^{-20} - 10^{-21}$
with constant frequency in the interval of 0,1 - 10 sec$^{-1}$.
Due to the expecting of the very small amplitude
of the gravitational radiation, the experimental
programs, oriented on direct gravitational radiation detection, run into
the problem of noise of shot and quantum character.
This contradiction has continued for 30 years and demands the
creation of high-precision deeply cooled detectors.

 On the other hand it is well-known that, even such weak by amplitude GW
consist of rather high energy  -  for the above examples this energy
is of the order of w/cm$^2$ for the first case and it is of the
order of $10^{-13} - 10^{-11}$ w/cm$^2$ for the second case. The registration
of the electromagnetic signal of that strength gives no problems.
Therefore the problem of the gravitational radiation detection should be
studied differently - to look for the  specific electromagnetic
signals as a result of the GW effect on the matter of those regions of Galaxy
where the gravitational radiation intensity is large .
Setting the problem this way we, first of all, study the effect of
gravitational waves on plasmalike media. The corresponding
investigations were fulfiled in the 80s years generally in the Kazan
school of Gravitation, and a number of specific electromagnetic
reflections of a plasma for plane gravitational waves (PGW) was revealed.
In \cite{zetp} - \cite{aniz2} the effect of PGW on plasmalike media was
investigated by the methods of relativistic kinetic theory in the
approximation when the back response of matter on the PGW is negligible :
\begin{equation}
\label{1.1} 8 \pi \varepsilon \ll \omega^{2},
\end{equation}
where $\omega$ is the characteristic PGW frequency, $\varepsilon$ is
the matter energy density ($G= \hbar =c= 1$). These papers have
revealed a number of phenomena of interest, consisting in the induction
of longitudinal electric oscillations in the plasma by PGW. In
spite of the strictness of the results obtained in [2-5], the effects
discovered in these
papers have very little to do with the real problem of GW detection.
Moreover, the above calculations show lack of any prospect for
GW detectors based on dynamic excitation of electric oscillations by
gravitational radiation. There are two reasons for that: the smallness of
the ratio $( m^{2} G/e^{2} ) = 10^{-43}$ and the small relativistic factor $
\langle v^{2} \rangle / c^{2}$ of standard plasmalike systems. The GW
energy transformation coefficient to plasma
oscillations is directly proportional to a product of these factors.

However, the situation may change radically, if strong electric
or magnetic fields are present in the plasma. In \cite{phlet}
where the induction of surface currents on the metal-vacuum interface by a
PGW was studied, it was shown that the values of currents thus induced
can be of experimental interest.
In \cite{probl} on the basis of relativistic kinetic equastions, a set of
magnetohydrodynamics (MHD) equations was obtained, which described
the motion of collisionless magnetoactive plasma on the
background of a PGW of an arbitrary magnitude  and it was shown that,
provided the propagation
of the PGW is transversal, there arises a plasma drift in the PGW
propagation direction.

\section{GMSW}

In Ref.\cite{gmsw} the exact solution of
relativistic MHD on the PGW background of an arbitrary magnitude was
obtained. On the basis of this a new class, that is of the
sufficiently nonlinear threshold effects, named GMSW  - gravitational magnetic
shock waves, was discovered.

PGW metrics of polarisation ${\bf e}_{+}$ are described by the
expression \cite{torn}:
\begin{equation}
\label{4.1} d s^{2}=2 du dv - L^{2}[e^{2 \beta}(dx^{2})^{2} +
e^{-2\beta}( dx^{3})^{2}],
\end{equation}
where $\beta(u)$ is an arbitrary function (the PGW amplitude);
the function $L(u)$ (the PGW background factor ) obeys an ardinary second
order differential equastion ;
$ u= \frac{1}{\sqrt{2}}(t - x^{1})$ is the retarted time
and $v= \frac{1}{\sqrt{2}}(t + x^{1})$ is the advanced time.
Let there be no PGW at ($u \leq 0$) :
\begin{equation}
\label{4.3} \beta(u)_{\mid u \leq 0}=0; \hspace{1 cm}
L(u)_{\mid u \leq 0}=1,
\end{equation}
the plasma is homogeneous and at rest:
\begin{displaymath}
v^{v}_{\mid u \leq 0}= v^{u}_{\mid u \leq 0}= \frac{1}{\sqrt{2}};
\hspace{1 cm} v^{2}=v^{3}=0;
\end{displaymath}
\begin{equation}
\label{4.4} \varepsilon_{\mid u \leq 0}=\varepsilon_{0};
\hspace{1cm} p_{\mid u \leq 0}=p_{0},
\end{equation}
($p = p(\varepsilon)$ is the pressure of the plasma, $v_k $ is the vector
of its dynamic velocity),
and a homogeneous magnetic field belongs to the plane
$\lbrace x^{1},x^{2} \rbrace $:
\begin{displaymath}
H_{1 \mid u \leq 0}=H_{0} \cos\Omega ; \hspace{1 cm}
H_{2 \mid u \leq 0}=H_{0} \sin\Omega ;
\end{displaymath}
\begin{equation}
\label{4.5} H_{3 \mid u \leq 0}=0; \hspace{1 cm}
E_{i \mid u \leq 0}=0,
\end{equation}
where $\Omega$ is the angle between the axis $0x^{1} $(the PGW propagation
direction) and the direction of the magnetic field
${\bf H}$. The conditions (\ref{4.5}) agree with the vector potential
\begin{displaymath}
A_{v}=A_{u}=A_{2}=0;
\end{displaymath}
\begin{equation}
\label{4.6} A_{3}=H_{0} (x^{1} \sin\Omega - x^{2} \cos\Omega);
\hspace{1.5 cm} (u \leq 0).
\end{equation}

The exact solution  of the relativistic MHD equastions
on the metrics background  (\ref{4.1}) obtained in \cite{gmsw}
satisfies the initial conditions (\ref{4.3}) - (\ref{4.5}) and is
determined by the {\em governing function}:
\begin{equation}
\label{4.39} \Delta(u) =
1 - \alpha^{2} \left[e^{2\beta} - 1\right]\,,
\end{equation}
where $\alpha$ is {\em dimensionless parameter}:
\begin{equation}
\label{4.40} \alpha^{2}=
 \frac{H^{2}_{0}\sin^{2}\Omega}{4\pi(\varepsilon_{0} + p_{0})}\,.
\end{equation}
This solution consists of a physical singularity
on the hyperserface $\Sigma_* :
u = u_*$:
\begin{equation}
\label{5.8}  \Delta(u_*) = 1 - \alpha^{2} \left[e^{2\beta(u_*)}
 - 1\right] = 0\,,
\end{equation}
on which the densities of the plasma energy and of the magnetic field
tend to infinity, the dynamic velocity of the plasma as a whole
tends to the velocity of light in the PGW propagation direction.
In this case  nd the relation of the magnetic field
energy density to the plasma energy tends to infinity.
The above singularity is
the GMSW spreading in the PGW propagation direction at a subluminal velocity.
In \cite{gmsw} it is shown that, in the frame of reference of an
external observer,
the particles are reaching the singular surface (\ref{5.8}) for
infinitely long time. According to Eq.(\ref{5.8}) the conditions of
the singularity arising are
\begin{equation}
\label{5.9} \beta(u) > 0;
\end{equation}
\begin{equation}
\label{5.10} \alpha^{2} > 1.
\end{equation}
The extremely important fact is that, the singular condition is even
possible in a weak PGW
$( | \beta| \ll 1)$ on the condition of a highly magnetized plasma
$(\alpha^{2} \gg 1)$. In this case the singular condition,
according to (\ref{5.8}), arises on the hypersurfaces $ u=u_{\ast}$:
\begin{equation}
\label{5.11} \beta(u_{\ast}) =  \frac{1}{2 \alpha^{2}}.
\end{equation}
In particular, in case of a barotropic equation of state
($ p = k \varepsilon , 0 \leq k < 1$):
\begin{equation}
\label{5.28} \varepsilon = \varepsilon_{0}
\Lambda^{-1- \nu} ;
\end{equation}
\begin{equation}
\label{5.29} v_{v} = \frac{1}{\sqrt{2}}
L^{\nu} \Delta^{1 + \nu/2} ;
\end{equation}
\begin{equation}
\label{5.30} \frac{v_{u}}{v_{v}} =
\Delta^{-2} \left[ \Lambda^{-\nu} + a \Delta -1)^{2} L^{-2} e^{-2\beta} \cot^{2}\Omega \right] ;
\end{equation}
\begin{equation}
\label{5.31} H^{2} = \frac{H^{2}_{0}}{\Lambda^{2}}
= \left( \cos^{2}\Omega + L^{2} \Lambda^{-\nu}
e^{2\beta} \sin^{2}\Omega \right) ,
\end{equation}
where  $\Lambda = L^{2}(u) \Delta(u)$ , $\nu = 2k/(1-k) \geq 0 $,
$H^2 = 1/2 F_{ik} F^{ik}$ is an invariant of the electromagnetic field,
(strength square of the magnetic field in the frame of reference
comeoving with the plasma).

It follows from (\ref{5.28}) - (\ref{5.31}) that, for $\beta > 0$ the plasma
moves in the GW  propagation direction ($v^1 = \frac{1}{\sqrt{2}}
(v_u - v_v) > 0$), by $\beta < 0$ - in the opposite direction.
The effect is a maximum in the PGW propagation direction, that is
perpendicular to the intensity of the original magnetic field, and
vanishes in the direction parallel to the magnetic field intensity.

The singular character of the solution (\ref{5.28}) - (\ref{5.31})
is the consequence of the approach, in which the magnetoactive
plasma has been described {\it on the background of vacuum} PGW. In this
case the gravitational wave plays the role of an infinite reservoir of energy.
At the surface (\ref{5.8}) the condition of the vacuum approximation
(\ref{1.1}) is broken. Therefor for correct describing the plasma
motion at the surface (\ref{5.8}) it is necessary to take into account the
back plasma influence on the PGW metrics.It is noted (\ref{gmsw}) that,
this leads to a strong (almost 100 \% )
absorbing of the GW the accellerated up to sublight velocity by plasma.

The energy of the shock wave is taken from the GW energy, thus
the GMSW is being an effective mechanism
\footnote{And as far as the author knows,
it is the only of all today-known mechanisms.}
transforming the GW energy into other types of energy
(generally into electromagnetic energy).
In \cite{rewiev} on the basis of the plasma energy and PGW balance a
self-consistent model is suggested.
It turned out that this process consists of a number of
common regularities \cite{rewiev}, \cite{vest}:
\begin{enumerate}
\item
Calculation of the plusma back influence on the GW removes the
singularity pointed above.
\item
GMSW is completely described by three nonnegative dimentionless parameters:
by the equation of plasma condition (parameter $k$), by the first
($\xi^2$):
\begin{equation} \label{xi}
\xi^2 \equiv \frac{\pi G \varepsilon_0 (1 + \alpha^2)}
{c^{2} \beta^2_0 \omega^{2}}
\end{equation}
and the second ($\Upsilon$) parameters of GMSW:
\begin{equation} \label{Upsilon}
\Upsilon = 2 \alpha^2 \beta_0\,,
\end{equation}
where $\beta_0$ is the maximum amplitude of vacuum PGW;
\item
The necessary condition of GMSW exitation is (\ref{5.9}) and (\ref{5.10}):
\begin{equation} \label{nessasary}
\Upsilon \geq 1\,;
\end{equation}
\item
The only criterion of the GW strong damping is the large value of the
second parameter of GMSW, ($\Upsilon$):
\begin{equation} \label{gg}
\Upsilon \gg 1\,;
\end{equation}
\item
In this case the plasma maximum response to GW\footnote{Here and further
speaking about the maximum response of plasma we mean its energetic
characteristics: the dencity of the plasma energy current and the dencity
of the magnetic field energy.} is achieved when the value of the first
parameter of  GMSW is small:
\begin{equation} \label{maxotklik}
\xi^2  \ll 1\,,
\end{equation}
\item
The plasma response to GW is like a single impulse. Moreover the shock
wave stage is always changed by the reverse stage, when the plasma turns
back. Simultaneously its density pressure and the magnetic field strength
damp;
\item
The plasma response, with the ultrarelativistic equation of state
($k = 1/3$), is much larger than the plasma
response with the nonrelativistic equation of state
($k = 0$), and in this case in an ultrarelativistic plasma the impulse
duration is a little bit larger;
\item
The GMSW impulse duration, $\tau$, does not exceed one forth of the GW period:
\begin{equation} \label{period}
\tau \leq \frac{T}{4} = \frac{\pi}{2 \omega}\,.
\end{equation}
\item
At a maximum of the GMSW impulse the density of the magnetic energy
reaches the value:
\begin{equation}  \label{maxotclik} \displaystyle
\left(\frac{H^2}{8 \pi}\right)_{max} = \frac{H_0^2}{8 \pi}
\frac{\sqrt{1 + \xi^2}}{xi}\,.
\end{equation}
\item
The local flash of the magnetic field intensity in the GMSW leads to
the power flash of the plasma magnetobremsstrahlung radiation,
wich is proportional to the magnetic field intensity square \cite{land}.
\end{enumerate}

\section{GMSW in the magne\-to\-spheres of neutron stars}
The only possible source of a GMSW can apparently be neutron stars
magnetospheres on the stage of the quadrupole oscillations of a neutron
star or on the stage of a Supernova \cite{gmsw}.

On the Fig.1. the relation of the GMSW first and second parameters in
the magnetosphere of the neutron star which is of the mass of
$M = 1,67 M_{\odot}$ (pulsar NP 0532) to the distance to the star center
is shown.

\ris{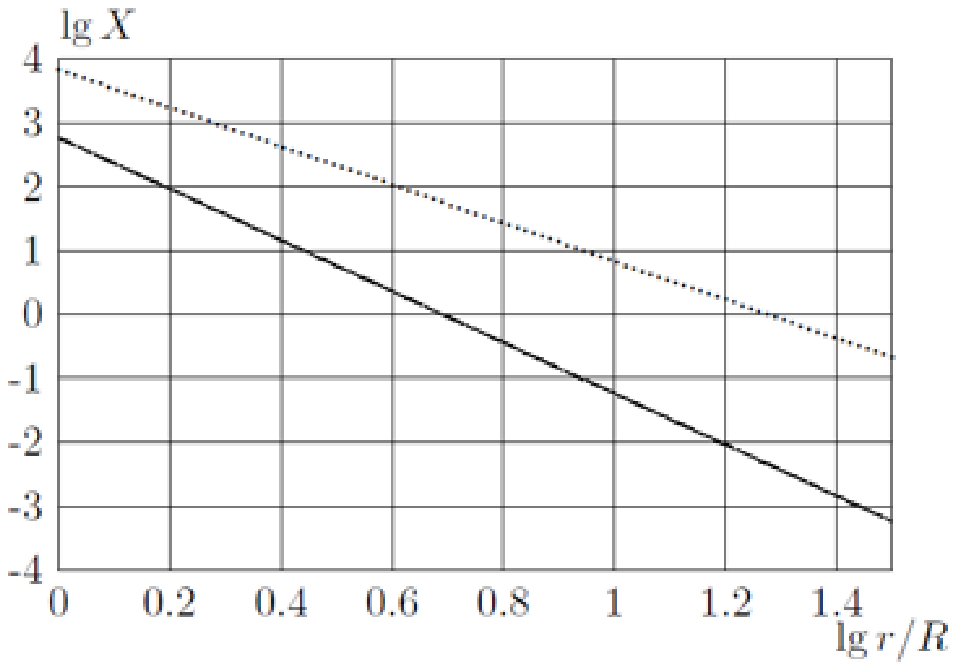}{GMSW parameters in a NP 0532 magnetosphere for a
dipole radial dependence of the magnetic field strength
$H\sim1/r$. Solid line: $\lg \xi^2$; dotted line: $\lg\Upsilon$.
We supposed $R=1.2\times10^6$cm in the neutron star radius,
$H(R)=10^{12}$G, $\beta_0(R)=10^{-8}$.}

In this case the electrons density in the magnetosphere was calculated
according to the estimate suggested in
\cite{pachini}:
\begin{equation} \label{ne} \displaystyle
n_e(R) \sim \frac{H}{4 \pi e R}\,.
\end{equation}
As we can see on this drawing in the region of $0.8 \leq \lg(r/R) \leq
1.2$, i.¥.,
$7.6 \cdot10^7 \leq r \leq 1.9 \cdot10^8$cm the conditions of the  GMSW
exitation are realized.
If the magnetic field of a neutron star is described as that of a dipole,
then the geographic angle $\Theta$ (counted relative to the magnetic
equator) will be connected with the above angle
$\Omega$ by the relation $\Omega = \pi/2 - \Theta $.
Therefore the GMSW excitation condition depends on the angle $\Theta$:
\begin{displaymath}
\label{7.1} \sin^{2}\Theta < 1 - \frac{1}{2\alpha^{2}_{0}|\beta|}.
\end{displaymath}

'hus, in the magnetosphere of a neutron star (or a Supernova ) a GMSW can
be excited in the region of the magnetic equator, similarly to pulsars
with a knife-like radiation pattern. In this region, as was demonstrated
by the above examples, the gravitational radiation can be absorbed
practically completely by the excitation of shock waves.
A neutron star of such type should radiate gravitational waves only from
its magnetic poles, similarly to pulsars with a pencil-like radiation
pattern. In this case the probability of observing gravitational sources
directly detecting a GW can be sharply dropped.

However the GMSW open another way of observing gravitational waves.
 The excited GMSW first of all carries with itself strong magnetic fields.
 These magnetic fields go from a neutron star in the magnetic equator
plane and therefore they have to encrease the intensity of the brake
radiation of the pulsar at the instant when the GMSW front is passing by.
Thus, at
the moment of the quadrupole oscillations exciting the anomalous intensity
splashes in the pulsar radiation must be observed. According to
(\ref{period})  duration of these flashes is to be shorter or of the
order of the GW period quarter and hence - of the quarter of the neutron star
own oscillations period.

According to \cite{torn} the frequency of neutron stars own oscillations
of neutron stars as a function of the mass of the latter variates in the
bounds of
$5\cdot 10^{3} \div 2\cdot 10^4$ sec$^{-1}$
mass. Therefore according to
(\ref{period}) the duration of the GMSW impulses and at the same time
of the radiation flashes must satisfy the condition:
\begin{equation}
\label{impuls} \tau < 3\cdot 10^{-4} \div 4\cdot 10^{-5} sec.
\end{equation}

Naturally the question about the nature of the pulsars quadrupole
oscillations araises. The nuclear responses of an explosion character with
heavy hyperons like
$n + n \longleftrightarrow p + \Sigma_{-}$ , that proceed in the nucleus
of neutron stars at densities more than
$10^{15}$ g/cm$^3$ \cite{lange} may be the energy source for those
oscillations. The presence of the strong magnetic field is to reduce to
the ansymmetry of the explosions, i.¥., to the quadrupole moment
excitation. A neutron star has to be rather young for such processes to happen
in it.
The calculations show \cite{camerun}, that after the Supernova explosion
the neutron star temperature falls approximately in an order during
$10^4$ years. Consequently, the GMSW is to be looked for in the
radiation of the rather young pulsars excited not earlier than 10,000 years
ago.
\section{Pulsar in Crabe NP 0532}
The pulsar with the required parametres exists - it is the famous
pusar in the Crabe nebula, NP 0532; its life-time is less than 1000
years (Supernova, 1054 year). This pulsar is the youngest
of all known ones (and, consequently, it is the hotest);
it has the shortest period: $T = 0,033$ sec.  But the most suprising fact
is that, in the radio-radiation of this pulsar one observes
anomals, that can be, in a great degree of assurance, indentified with the
GMSW. That is: in the radioradiation NP 0532 one observes single
non-regular , so-called {\em giant impulses} (in the average
1 impulse an every 5 - 10 minutes) \cite{pulsar}. The intensity
of the radiation by the giant impulses is larger in tens times (approximately in 60 times),
than in common impulses.
But the most interesting fact is that, the duration of the giant impulses
is not more than $9 \cdot 10^{-5}$ sec, i.¥., almost in 2(!)
orders shorter than that of the usual pulsars
NP 0532 ($\tau \sim 6\cdot 10^{-3}$ sec ).
The duration of the usual impulses, as it is not difficult to see,
is of the order of 1/5 of the rotation period. 'hus, the common
pulsars are clearly geometrically explained by the pulsar rotation. The
duration of the giant impulses is 300 times shorter than the rotation
NP 0532 period, in consequence of this fact there is still no satisfying
theoretical model for the excitation of the giant impulses.

However, the giant impulses can be easily explained as the GMSW,
while the duration of the giant impulses is related not to the period
 of rotation of the pulsar and the angle of the knife-like pattern
of the radiation direction , but to the period of its own oscillations.
The comparing of the durations of the giant NP 0532 impulses
with the duration of the GMSW impulse (\ref{impuls})
shows a striking coincidence of these values. Really, in case of the NP 0532
pulsar bythe annilation line shift in the spectrum of $\gamma$ - radiation
(400 kev instead of 511) the gravitational red displacement is known
\cite{varma} and \cite{surface}: $\Delta E/E = $ $M G/R c^2 = 0.217\,.$
The extrapolation of the neutron star calculations \cite{torn}
gives the NP 0532 pulsar mass: $M=1.67 M_{\odot}$
and the radius of the corresponding neutron star: $R = 12$km.
According to the estimates \cite{pulsar} the magnetic field strength of
the NP 0532 pulsar surface is of $10^{12}$G.
Then according to (\ref{period}) the GMSW impulse duration for the
NP 0532 pulsar is to be approximately 87 microseconds. The
observed duration of the giant NP 0532 impulse is approximately
90 microseconds \cite{pulsar} (!).
For the explanation of the observed radiation the
amplitude maximum GW values on the stellar surface of the order of
$10^{-8}$ are requied. Such an amplitude corresponds to the gravitational
radiation power of the order of $4\cdot10^{42}$ erg/sec and the neutron
star oscillations energy of the order of $E_m \approx 4\cdot10^{41}$erg.
Taking into account that for the lifetime the NP 0532 pulsar of
(1000 years) approximately $7\cdot10^7$ giant impulses have been radiated,
we we get the estimate of the energy which was brought from the neutron
star by the gravitational waves
$E = 2.8\cdot10^{49}$erg. It is $10^{-5}$ of the rest energy of the given
neutron star This is completely connected with the suggestion about
the constant rebuilding of its nucleus.

Note that, only the PSR 0833 of all known pulsars, apparently can radiate
(but more seldom ) the giant impulses . The other known pulsars are
 too old to do that . Therefore it is necessary to focus
on the observing of these pulsars.
It is worth to especially emphasize that, there are no other mechanisms
which can accelerate the shock wave till the sublight
velocities . Therefore it is very important to study
the spectrum of the giant impulses in the X-ray range
with the aim of discovering the violet shift of the radiation spectrum .
Detailed studing of the giant impulses (their forms and moment spectrum)
will allow not only to verify the existence of
gravitational radiation , but
also to give additional information about neutron star structure and
about the processes taking place in their interior.
In turn it is necessary to research the GMSW-impulses formation in detail
theoretically. We intend to study this problem in future papers.

\end{document}